# Refraction, the speed of light and minimal action:

# From Descartes to Maupertuis through many more

Shahen Hacyan[1]

Instituto de Física,
Universidad Nacional Autónoma de México
Mexico City, Mexico

**Abstract:** In the 17th and 18th centuries, several natural philosophers studied the phenomenon of refraction and attempted to obtain the Snell law from various assumptions. Lacking experimental data, it was generally believed that light travels faster in a refracting medium than in air. In the present article, I review the contributions to the problem of light refraction by Descartes, Fermat, Huygens, Leibniz, Newton, Clairaut, and finally Maupertuis who established a principle of least action based on his own approach to the problem.

1. Introduction

The refraction of light and what is known as Snell's law (i.e., the angular relationship between incident and refracted rays) has a long history that goes back to the ancient Greek and Arab philosophers (see, e.g., Darrigol (2012) for a complete history). During the 17th and 18th centuries, many distinguished philosophers tried to explain refraction in terms of physical processes and mathematical calculations. It should be noted that at such a time, the speed of light in a medium was not known. Many influential scientists who studied refraction, such as Descartes and Newton, believed that this speed is higher in a transparent medium than in air; they tried to justify the Snell law accordingly. Fermat, on the contrary, derived Snell´s law on the assumption that light takes the minimum time to travel from one point to another and that its speed is slower in a denser medium; as expected, his work received fierce criticisms from the scientific establishment of his time, dominated in France by the Cartesians (Degas, 1950; Hacyan, 2023). A correct derivation (by present days standards) of Snell's law was given by Huygens in 1690 in the context of his own theory of light as a wave. He postulated that the waves of light expand more slowly in a medium than in air, thus changing their direction of propagation. Huygens's theory, however, had to compete with Newton's theory of light as made up of particles; in the case of refraction, Newton believed that the speed of light is increased by some force inside the medium.

In a rather overlooked article, Leibniz derived Snell's law using an approach like Fermat's, but looking for the "easiest" path rather than the fastest. In this context, the subsequent contribution of Maupertuis is particularly relevant: he also assumed that the speed of light is higher in a medium, but he associated the problem of refraction to a much more general law, the principle of minimal action, which he elevated to the status of a fundamental law of Nature (Maupertuis, 1744 and 1746).

---

[1] hacyan@fisica.unam.mx

Of course, it would be rash to underestimate the work of these early scientists for relying on a wrong assumption on the propagation of light. It must be noted that the speed of light in their time was only known from astronomical observations. It was not at all obvious that light should move more slowly in a medium, since it was already known that sound, on the contrary, propagates faster in a denser medium; for example, the speed of sound in water is more than four times its speed in air, and sound is even faster in metals due to their stiffness. Thus, there was no reason to expect anything different from light.

Actually, the first measurement of the speed of light in a terrestrial laboratory was performed by Hippolyte Fizeau (1819 - 1896) in 1849. Two years later, Fizeau used his experimental setup to measure the speed of light in water –stagnant or moving– and established beyond any doubt that light moves more slowly in water than in air, thus ending a time-long discussion (Darrigol, 2012).

In the present paper, I will review the contributions of Descartes, Fermat, Huygens, Leibniz, Newton, Clairaut, and Maupertuis to the problem of the refraction of light, with some emphasis on the principle of minimal action, which was outlined by Fermat and Leibniz, and given a more general form by Maupertuis.

## 2. Descartes

René Descartes (1596 – 1650) published in 1637 the *Dioptrique*, a treatise on light which also included a study of the human eye and the techniques for polishing lenses. In the chapter devoted to the refraction of light, he first considered, as a warm-up exercise, the case of a ball thrown onto a solid surface or across a piece of cloth, and the way its trajectory is deviated due to the change of speed in each case. He then considered the case of a light ray penetrating a transparent medium such as water. The key to his reasoning can be seen from the figure in his text (here, Fig. 1). Without water, the ray would follow a straight trajectory *ABD*, but due to the presence of water, the speed of light in the direction parallel to the surface *CBE* increases and thus light traverses the distance *BE* in less time than if there were no water. By identifying this enhanced velocity with the refraction index, Descartes obtained the Snell law with some geometrical manipulations (even though, he did not summarize his result in terms of sines).

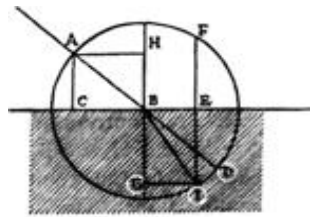

Fig. 1

In case there were any doubt about the speed of light in a medium, Descartes attempted to prove his assumption based on his own speculations on the nature of light and matter. According to his theory, light was the motion of a "very subtle matter" (*une matière très subtile*) which filled the Universe. Light must move faster in a medium, he claimed, because "the harder and firmer the small parts of a transparent body are, the more they allow light to pass easily through because this light does not need to remove none [of these small parts] from their positions".[2]

Descartes' speculations were taken most seriously by the next generations of French philosophers. His theories of matter and light, and refraction, were well accepted, albeit a better approach (as we presently know) was due to Fermat.

### 3. Fermat

In an article published in 1662 (see, e.g., Dugas, 1950, 1988), Pierre Fermat (1601-1665) proposed another method to get the law of refraction. Suppose a particle moves from point *M* in a medium to point *N* in another medium (Fig. 2); the problem is to find the trajectory that takes the least *time* if the speed in each medium is different. Fermat proved that such a trajectory satisfies the law of sines if the speed in the first medium (say, air) is higher than in the second (say, water).

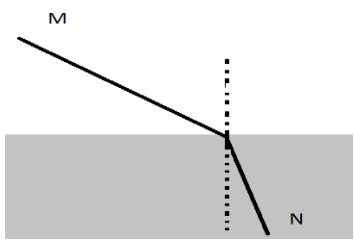

Fig. 2

Fermat's basic assumption, that light travels *more slowly* in a medium than in air, was against his Cartesian colleagues. They harshly criticized him for contradicting their authority, but also for relying on purely mathematical arguments to explain a physical process (Dugas, 1950; Hacyan, 2023). Accordingly, the problem of finding the trajectory that takes the least time was dismissed by Fermat's contemporaries. However, the basic idea that something should be minimized remained implicit, to become explicit less than a century later.

### 4. Huygens

Christiaan Huygens (1629-1695) presented his theory of light in the extensive *Traité de la Lumière* (Treatise on Light), which appeared in 1690. According to his theory, light was the vibration of a very subtle substance, the Ether, that filled the whole Universe. Being a wave, light should be analogous to sound in the air or waves in water, and likewise propagate spherically from its source. As for the trajectory of light in a medium, Huygens speculated that the transparency of certain materials might be due to the penetration of the Ether between the small particles tied together, which constituted the material medium. In his own words:

> We easily conceive that waves could continue in the ethereal material that fills the interstices of the particles. And moreover, we can believe that the progress of these waves must be somewhat slower inside the bodies, due to the small detours caused by these same particles. I will show such a difference of velocities to be the cause of refraction.

---

[2] All translations are mine.

Thus, with the assumption that spherical waves propagate more slowly in a transparent medium, Huygens derived the law of sines in a clear geometrical way. His deduction is perfectly valid and consistent with his assumptions.

5. **Leibniz**

Inspired by Fermat's treatment of refraction, Gottfried Wilhelm Leibniz (1646 – 1716) made an important contribution to Optics in *Tentamen Anagogicum* (Anagogical Essay)*,* an essay written in 1696 and published posthumously, largely devoted to the existence of "final causes*".* To support his philosophical point of view, Leibniz claimed that Nature always acts with the purpose of using the "easiest way of all" (*viâ omnium facillimâ*) and he illustrated his point by attacking the problem of refraction as Fermat had done. However, unlike Fermat, he did not look for the path taken by light in the shortest time, but for the "easiest" of all paths. What he meant by "easiest" was the minimum value that could take the product of the *distance* traveled by light and the *resistance* of the medium to its passage. Of course, it is not obvious how to determine such a resistance, but for all practical purposes, it was enough to take it as a free parameter.

Leibniz's calculations are rather confusing, but they can be summarized with a modern notation and a simplified version of the figures included in his article (see McDonough 2009 for a detailed account of the whole article). The scheme proposed by Leibniz is basically as synthesized in Figure 3 below: a ray starts in point *F* in one medium and ends up at point *G* in another medium, breaking its straight path at point *C*. According to Leibniz, the easiest path is the one for which the quantity

$$f\ CF + g\ CG \qquad (1)$$

takes a minimum value, with *f* and *g* being the resistances of the media of *F* and *G* respectively. Taking the differential of the above quantity and equating it to zero in order to find its minimum, Leibniz obtained the relation

$$\frac{CF}{CG} = \frac{f\ PF}{g\ PG},$$

where the point *P* is the intersection of the straight-line *FG* with the perpendicular to the surface drawn from *C* (see Fig. 3). Then, Leibniz showed with some simple geometry that

$$\frac{\sin i}{\sin r} = \frac{g}{f}, \qquad (2)$$

where *i* and *r* are the incidence and refraction angles, respectively, as seen in Fig. 3. As an illustrative example, suppose that *F* is located in vacuum (or air) and *G* inside a transparent medium. Since it is well known that in this case the incidence angle *i* is higher than the refraction angle *r*, it follows that the resistance *g* of the medium must be higher than the resistance *f* in vacuum (or air). This sounds reasonable, except that Leibniz believed, like most of his colleagues, that the speed of light is higher in a medium. To solve this apparent contradiction, he previously had proposed in a short article of 1682 that a greater resistance hinders the diffusion of light and hence must increase its speed.

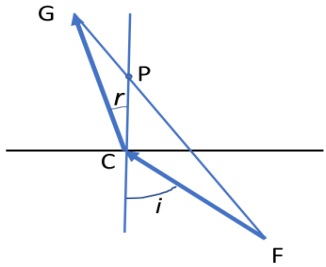

Fig. 3

Let us compare Leibniz's treatment of the problem with that of Fermat's, for whom time —which is distance *divided* by speed— was the quantity to be minimized. Accordingly, instead of expression (1), Fermat looked for the minimum of

$$\frac{CF}{v_f} + \frac{CG}{v_g}, \qquad (3)$$

where $v_f$ and $v_g$ are the speeds of light in the media of F and G, respectively. If F is in vacuum, then $v_f = c$ (the speed of light) and $v_g = c/n$ (n is the refraction index). It then follows that

$$\frac{\sin i}{\sin r} = n \qquad (4)$$

and n > 1, as it should be.

Summing up, we see that, from a mathematical point of view, Fermat and Leibniz followed a similar path, although the latter had the calculus of maxima and minima (he himself had coinvented) to his great advantage. It also follows from the above approach that what Leibniz called *resistance* should be, by today's standards, inversely proportional to the speed of light in the medium.

6. **Newton**

In 1704, Isaac Newton (1642-1627) finally published the *Opticks,* his classical treatise on light (just after the death of his nemesis Robert Hooke, but that is another story (see Westfall, 1981)). Contrary to Huygens, Newton believed that light is made up of particles that propagate freely in space and also through some materials. In the chapter devoted to refraction, Newton presented his own explanation of this phenomenon (Proposition X, Book Two, Part III):

> If light be swifter in bodies than in vacuo, in the proportion of the Sines which measure the refraction of the bodies, the forces of the bodies to reflect and refract light, are very nearly proportional to the densities of the same bodies…

His basic assumption was that there were forces in a refracting medium directed perpendicularly to their parallel surfaces, such that they caused the light passing through the medium to move faster. He calculated in a table the "refractive forces" derived directly from the law of sines for various refracting bodies: air, glass, topaz, oil, rainwater, etc. However, he never specified the nature of such forces inside the media. He only concluded that "all bodies seem to have their refractive powers proportional to their densities (or very nearly)", a power that may be due, he conjectured, to "sulphureous oily particles" which are more or less present in all bodies. Clearly, the argument was not quite convincing, especially compared to Huygens 'explanation of the same phenomenon.

7. Clairaut

Alexis Claude Clairaut (1713-1765) was an outstanding disciple of Maupertuis (whom we will meet in the next section) who followed his teacher in defending the physics of Newton in France. In an article published in 1739, Clairaut endorsed Newton's assumption that the phenomenon of refraction is due to a certain force acting on the particles of light. However, unlike Newton who believed that such a force is present inside a transparent medium, Clairaut conjectured that the deflecting force is constrained to a very thin "atmosphere" covering the surfaces of the refracting medium on both sides. Thus, according to Clairaut´s supposition, a ray would be deflected when entering such an "atmosphere" and then deflected again when leaving from the other side (see Fig. 4). The speed of light according to this scheme is not altered inside the medium, but only in the "atmosphere" that accelerates it. Thus, with some tedious algebra, Clairaut managed to get the law of sines from his calculation of the trajectory of a particle passing through a medium, although his theory did not stand the test of time.

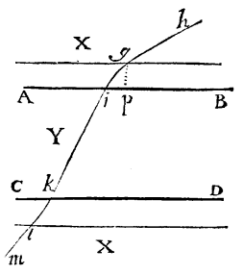

Fig. 4

8. Maupertuis

Pierre Moreau de Maupertuis (1698 – 1759) was one of the few early French scientists who championed the physics of Newton against the Cartesian establishment of his time (La Beaumelle, 1856; Terral, 2002; Shank, 2008). He went down in history for his expedition to Lapland (with Clairaut and others) to measure the meridian of the Earth and thus confirm that our planet has the shape of a flattened ellipsoid, just as Newton had predicted. He is also known for establishing a principle that bears his name: principle of least action. Its origin was Maupertuis' attempt to explain refraction, probably inspired by Fermat's work. However, unlike Fermat, Maupertuis believed that the speed of light was higher in a refracting medium than in air, as his fellow Cartesians. Accordingly, the physical quantity that should be minimized was not time, but something else, the action, that he defined as the product of velocity and distance.

In the figure that appeared in his scientific article presented in 1744 (here Fig. 5), a ray of light leaves the point *A* and reaches the point *B* inside the refracting medium. Supposing that *m* and *n* are the speeds of light in the upper and lower spaces respectively, the problem is to find the breaking point *R* such that the action

$$m\,AR + n\,RB$$

is a minimum. As a simple exercise in calculus, Maupertuis proved that the minimum of this quantity occurs, in all cases, if the sine of the incidence angle is to the sine of the refracted angle as *n/m*. From where he deduced that the speed *n* in the refracting medium must be *higher* than the speed *m* in air to have a refraction as depicted in the figure, since in this case the angle of incidence is bigger than the angle of refraction.

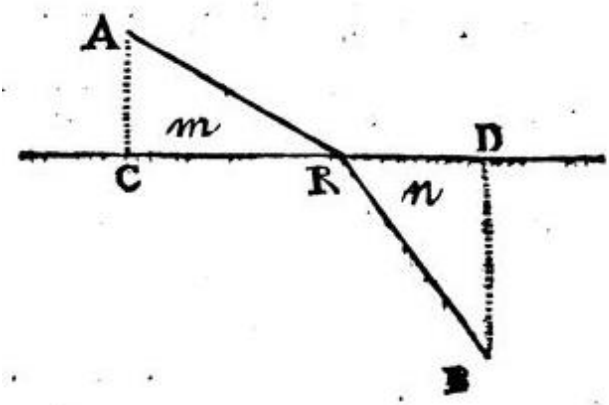

Fig. 5

Maupertuis was convinced that he had discovered a "metaphysical law" established by God: namely that "Nature, in the production of its effects, always acts by the *simplest* means". Thus, he realized that his principle of minimal action could cover more general cases than light.

In a second article, presented to the Berlin Royal Academy of Sciences in 1746, Maupertuis extended the definition of action to the product of mass times speed and distance. This is the quantity that must be a minimum. As an example, he applied this hypothesis to the (linear) collision of particles, either hard or elastic, and obtained the correct results for these problems.

In modern notation, Maupertuis' action is the integral

$$\int m\,v\,ds,$$

where *v* is the speed and *ds* a differential element of length along the trajectory of a particle. Of course, since *ds = v dt,* this action is the same as

$$\int m\,v^2\,dt,$$

where we recognize the famous *vis viva* or, in modern terms, the kinetic energy without the factor ½.

As a historical curiosity, it is worth noting that the principle of least action gave rise to a controversy over its paternity. Johann Samuel König (1712 – 1757), a colleague of Maupertuis and fan of Leibniz, claimed that such a principle was originally proposed by Leibniz, supposedly in a private letter… which was never found (La Beaumelle, 1856; Terral, 2002). The only relevant issue was that Leonhard Euler (1707 -1783) involved himself in the discussion. In a *Dissertatio* published in 1753, he agreed that many authors, from antiquity to his time, had assumed that Nature acts by the simplest ways, but he gave full credit to Maupertuis for having expressed this notion in a precise form. Leibniz had indeed considered the problem of refraction with the supposition that Nature always chooses to spend the less amount of what he defined as the product of the distance and the resistance of a medium. In any case, as Euler noted, Leibniz was in accordance with Descartes in supposing that light is faster in a medium than in air, notwithstanding that the medium has a higher resistance. To solve this apparent paradox, Euler agreed with Leibniz that a greater resistance must force light to move faster, and he gave the example of a fluid flowing faster through a narrower channel, a fact that may seem counterintuitive at first sight (it is due to Bernoulli's principle).

9. **Outcome**

In retrospective, Maupertuis' principle of minimal action was inspired by the wrong assumption about the speed of light in a medium, while Fermat's approach to the problem of refraction turned out to be correct by current standards. Nevertheless, the principle sketched out by Maupertuis was later established in a rigorous way by Euler and Joseph-Louis Lagrange (1736-1813). As it can be seen in modern textbooks on Mechanics, what is usually called the Maupertuis action is defined as $\int \boldsymbol{p} \cdot \boldsymbol{dq}$, where $\boldsymbol{q}$ and $\boldsymbol{p}$ are generalized coordinates and momenta. As for light, we now know that Maupertuis' principle, as originally proposed, does not apply to optical phenomena in the simple form he believed. The correct statement of this principle and its precise relationship to Optics was definitely established by Willian Rowan Hamilton (1805 -1865) a century later through the important principle that bears his name (see, e.g., Dugas, particularly Chapter VI).